\documentclass[english,aps,twocolumn]{revtex4}
\usepackage{epsfig}
\usepackage{graphicx}
\usepackage{amsmath}

\makeatletter
\@ifundefined{textcolor}{}
{%
 \definecolor{BLACK}{gray}{0}
 \definecolor{WHITE}{gray}{1}
 \definecolor{RED}{rgb}{1,0,0}
 \definecolor{GREEN}{rgb}{0,1,0}
 \definecolor{BLUE}{rgb}{0,0,1}
 \definecolor{CYAN}{cmyk}{1,0,0,0}
 \definecolor{MAGENTA}{cmyk}{0,1,0,0}
 \definecolor{YELLOW}{cmyk}{0,0,1,0}
 }

\makeatother

\begin{document}
\title{Nucleosynthesis: what direct reactions can do for it?%
\thanks{Presented at the Zakopane Conference on Nuclear Physics, Poland, 2012.}%
}
\author{Carlos A. Bertulani}
\affiliation{Department of Physics and Astronomy, Texas A\&M University-Commerce, Commerce,
TX 75429, USA\footnote{Email: carlos.bertulani@tamuc.edu}
}
\begin{abstract}
The reactions of relevance for stellar evolution are difficult to measure directly in the
laboratory at the small astrophysical energies. In recent years
indirect reaction methods have been developed and applied to extract
low-energy astrophysical S-factors. These methods require a
combination of new experimental techniques and theoretical efforts,
which are the subject of this review.
\end{abstract}

  
\maketitle

\section{Introduction}

{\bf Fusion}  - Fusion cross sections can be calculated from the equation \cite{Can06}
\begin{equation}
\sigma_F(E)=\pi {\lambdabar}^2 \sum_\ell (2\ell +1) P_\ell (E) , \label{eqnf}
\end{equation}
where $E$ is the center of mass energy, $\lambdabar=\sqrt{\hbar^2/2mE}$ is the reduced wavelength and $\ell=0,1,2,\cdots$.
The cross section is proportional to $\pi\lambdabar^2$, the area of the quantum wave. Each part of the wave corresponds to different impact parameters having different probabilities for fusion. As the impact parameter increases, so does the angular momentum, hence the reason for the $2\ell+1$ term. $P_\ell(E)$ is the probability that  fusion occurs at a given impact parameter, or angular momentum. 
Sometimes, for a better visualization, or for extrapolation to low energies, one uses the concept of {\it astrophysical S-factor}, redefining the cross section as   
\begin{equation}
\sigma_F(E)={1\over E} S(E)\exp\left[ -2\pi \eta(E)\right], \label{SE}
\end{equation} 
where $\eta(E)=Z_1Z_2e^2/\hbar v$, with $v$ being the relative velocity. The exponential function is an approximation to $P_0(E)$ for a square-well nuclear potential plus Coulomb potential, whereas the factor $1/E$ is proportional to the area appearing in Eq. \ref{eqnf}.   

{\bf Many reaction channels} - Eq. \ref{eqnf} does not work in most situations.  Only by including coupling to other channels, the fusion cross sections can be reproduced.  In  coupled channels schemes  one expands the total wavefunction for the system as
\begin{equation}
\Psi=\sum_{i} a_i(\alpha)\phi_i(\alpha, q_k),
\end{equation} 
where $\phi$ form the channel  basis, $\alpha$ is a dynamical variable (e.g., the distance between the nuclei), and $q_k$ are intrinsic coordinates. Inserting this expansion in the Schr\"odinger equation yields  a set of CC equations in the form \cite{Ber05}
\begin{equation}
{da_k\over d\alpha}=\sum_j a_j  \ \langle \phi_k \left| U \right| \phi_j \rangle \ e^{iE_\alpha \alpha/\hbar}, \label{cc}
\end{equation}
where $U$ is whatever potential couples the channels $k$ and $j$ and $E_\alpha=E_\alpha^{(k)}-E_\alpha^{(j)}$ is some sort of transition energy, or transition momentum. In the presence of continuum states, continuum-continuum coupling (relevant for breakup channels) can be included by discretizing the continuum. This goes by the name of {\it Continuum Discretized Coupled-Channels} (CDCC) calculations. There are several variations of CC equations, e.g., a set of differential equations for the wavefunctions, instead of using basis amplitudes. Coupled channels calculations with a large number of channels in continuum couplings are somewhat challenging:  the phases of matrix elements can add destructively or constructively, depending on the system and on the nuclear model. Such suppressions or enhancements are difficult to interpret. 

{\bf Radiative capture} - For reactions involving light nuclei, only a few channels are of relevance. In this case, a real potential  is enough for the treatment of fusion. For example,  radiative capture cross sections of the type $n+x\rightarrow a+\gamma$
and $\pi L$ ($\pi=E,(M)=$electric (magnetic) L-pole) transitions
can be calculated from \cite{Ber94}
\begin{equation}
\sigma_{EL,J_{b}}^{\rm d.c.}   = const. \times \left\vert \left\langle l_cj_c\left\Vert
\mathcal{O}_{\pi L}\right\Vert l_{b}j_{b} \right\rangle
\right\vert^{2},
\label{respf}%
\end{equation}
where $\mathcal{O}_{\pi L}$ is an EM operator, and $\left\langle
l_cj_c\left\Vert \mathcal{O}_{\pi L}\right\Vert l_{b}j_{b}
\right\rangle$ is a multipole matrix element involving bound ($b$) and continuum ($c$) wavefunctons. For 
electric multipole transitions ($ \mathcal{O}_{\pi L}= r^LY_{LM}$),
\begin{equation}
\left\langle l_cj_c\left\Vert
\mathcal{O}_{EL}\right\Vert l_{b} j_{b}\right\rangle
=const. \times \int_{0}^{\infty}dr \
r^{L}u_{b}(r)u_{c}(r)
,\label{lol0}%
\end{equation}
where $u_i$ are radial wavefunctions.
The total direct capture cross section is obtained by adding all
multipolarities and final spins of the bound state ($E\equiv E_{nx}$),
\begin{equation}
\sigma^{{\rm d.c.}} (E)=\sum_{L,J_{b}} (SF)_{J_{b}}\ \sigma^{{\rm d.c.}%
}_{L,J_{b}}(E) \ , \label{SFS}%
\end{equation}
where $(SF)_{J_{b}}$ are spectroscopic factors.

{\bf Asymptotic normalization coefficients} - In a microscopic approach,
instead of single-particle wavefunctions one often makes use of overlap
integrals, $I_{b}(r)$, and a many-body wavefunction for the relative
motion, $u_{c} (r)$. Both $I_{b}(r)$ and $u
_{c}(r)$ might be very complicated to calculate, depending on how
elaborated the microscopic model is. The variable $r$ is the relative
coordinate between the nucleon and the nucleus $x$, with all the intrinsic
coordinates of the nucleons in $x$ being integrated out. The direct capture
cross sections are obtained from the calculation of $\sigma_{L,J_{b}%
}^{{\rm d.c.}} \propto|\left<  I_{b}(r)||r^{L}Y_{L}|| \Psi_{c}(r)\right>
|^{2}$.

The imprints of many-body effects will eventually disappear at large distances
between the nucleon and the nucleus. One thus expects that the overlap
function asymptotically matches ($r\rightarrow\infty$),
\begin{eqnarray}
I_{b}(r)    =C_{i} \times \left( \frac{W_{-\eta,l_{b}+1/2}(2\kappa r)}{r} \right) \  \left[
\sqrt{\frac{2\kappa}{r}}K_{l_{b}+1/2}(\kappa r) \right] , \label{whitt}%
\end{eqnarray}
where () are for protons and [] for neutrons. The binding energy of the $n+x$ system is related to $\kappa$ by means
of $E_{b}=\hbar^{2}\kappa^{2}/2m_{nx}$, $W_{p,q}$ is the Whittaker function
and $K_{\mu}$ is the modified Bessel function. In      Eq. \ref{whitt}, $C_{i}$ is
the {\it Asymptotic Normalization Coefficient} (ANC).

In the calculation of $\sigma_{L,J_{b}}^{{\rm d.c.}}$ above, one often
meets the situation in which only the asymptotic part of $I_{b}(r)$ and
$\Psi_{c}(r)$ contributes significantly to the integral over $r$. In these
situations, $u_{c}(r)$ is also well described by a simple two-body
scattering wave (e.g. Coulomb waves). Therefore the radial integration in
$\sigma_{L,J_{b}}^{{\rm d.c.}}$ can be done accurately and the only
remaining information from the many-body physics at short-distances is
contained in the asymptotic normalization coefficient $C_{i}$, i.e.
$\sigma_{L,J_{b}}^{{\rm d.c.}}\propto C_{i}^{2}$. We thus run into an
{\it effective theory for radiative capture cross sections}, in which the constants
$C_{i}$ carry all the information about the short-distance physics, where the
many-body aspects are relevant \cite{Muk90} . It is worthwhile to mention that these
arguments are reasonable for proton capture at very low energies, because of
the Coulomb barrier.

{\bf Resonating group method} - One immediate goal can be achieved in the coming years by using the
{\it Resonating Group Method} (RGM) or the
Generator Coordinate Method (GCM). These are a set of coupled integro-differential
equations of the form \cite{TLT78}
\begin{equation}
\sum_{\alpha'} \int d^3 r'
\left[
H^{AB}_{\alpha\alpha'}({\bf r,r'})-EN^{AB}_{\alpha\alpha'}({\bf r,r'})
\right]
g_{\alpha'}({\bf r'})=0,\label{RGM}
\end{equation}
where $H(N)({\bf r,r'})=\langle \Psi_A(\alpha,{\bf r})|H(1)|
\Psi_B(\alpha',{\bf r'}) \rangle$. In these equations $H$ is the Hamiltonian for the
system of two nuclei (A and B) with the energy $E$, $\Psi_{A,B}$ is the wavefunction
of nucleus A (and B), and $g_{\alpha}({\bf r})$ is a function to be found by numerical
solution of Eq. \ref{RGM}, which describes the relative motion of A and B in channel
$\alpha$.
Full antisymmetrization between nucleons of A and B are implicit.
Modern nuclear shell-model calculations are able to
provide the wavefunctions $\Psi_{A,B}$ for light nuclei \cite{QN08}. 

\begin{figure}
\begin{center}\includegraphics[
height=3. in]{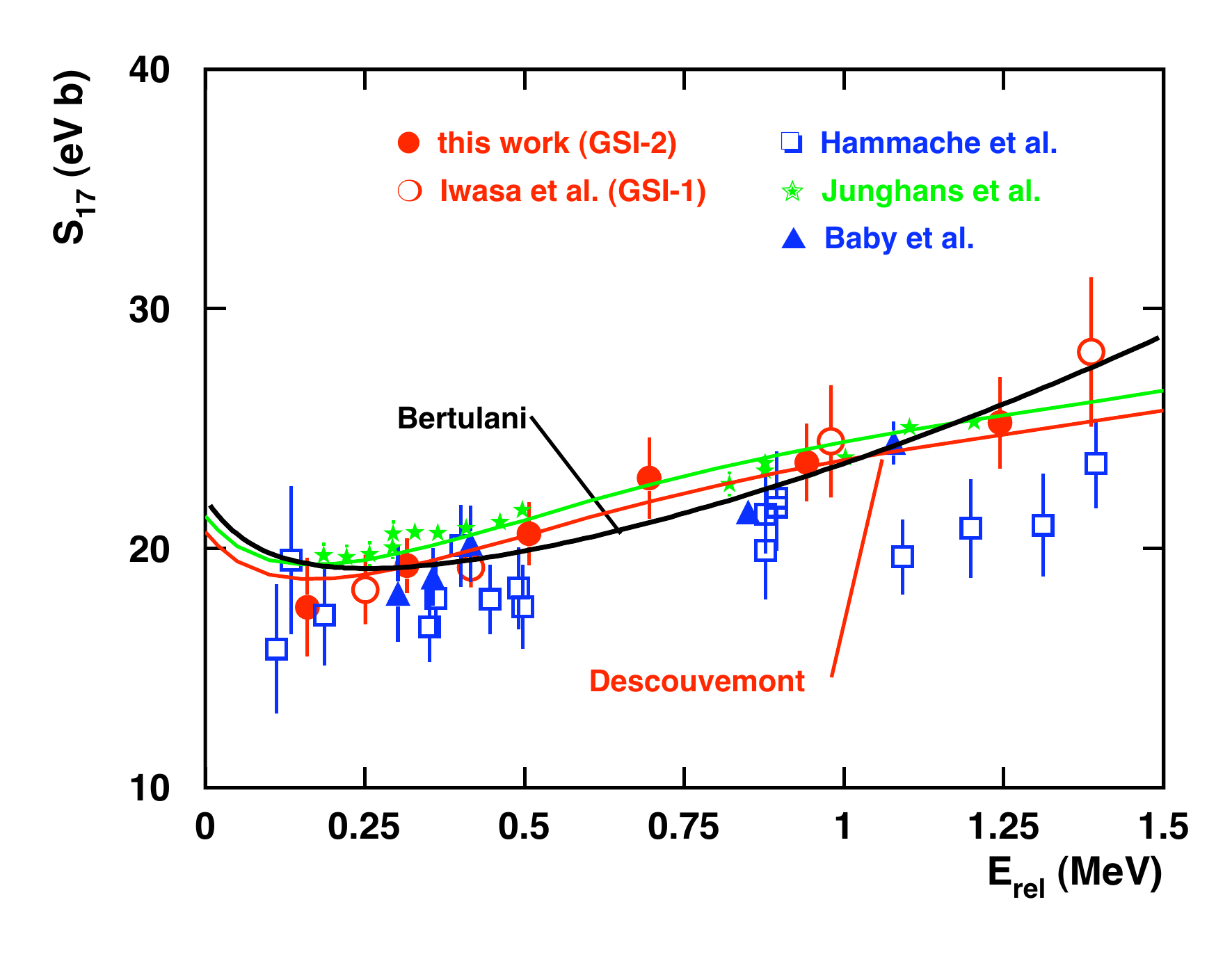}
\caption{ World data on $^7$Be(p,$\gamma$)$^8$B compared to theoretical calculations. }
\label{s17ber}
\end{center}
\end{figure}

The astrophysical S-factor
for the reaction  $^{7}$Be$($p$,\gamma)^{8}$B was  calculated \cite{NBC05}
and excellent agreement was found with the experimental data in both
direct and indirect measurements \cite{NBC05,PRQ11}. The low- and
high-energy slopes of the  S-factor obtained with a many-body microscopic calculation \cite{NBC05} is well
described by the fit 
\begin{equation}
S_{17}(E)=(22.109\ {\rm eV.b}){1+5.30E+1.65E^2+0.857E^3 \over 1+E/0.1375}  ,
\end{equation}
where E is the relative energy (in MeV) of p$+^7$Be in their
center-of-mass. This equation corresponds to a Pad\'e approximant of
the S-factor. A subthreshold pole due to the binding energy of $^8$B
is responsible for the denominator \cite{JKS98,WK81}. Figure \ref{s17ber} show the world data on $^7$Be(p,$\gamma$)$^8$B compared to a few of the theoretical calculations. The recent compilation published in Ref. \cite{RMP11} recommends $S_{17} = 20.8 \pm 0.7 \ {\rm (expt)} \pm 1.4\ {\rm (theor)}$ eV b.

\section{Direct reactions and the role of radioactive beams}

{\bf Transfer reactions} -  Transfer reactions $A(a,b)B$ are effective when a momentum matching
exists between the transferred particle and the internal particles
in the nucleus. Thus, beam energies should be in the range of a few
10 MeV per nucleon. Low energy reactions of
astrophysical interest can be extracted directly from breakup
reactions $A+a \longrightarrow b+c+B$ by means of the  {\it Trojan
Horse method} (THM)  \cite{Bau86}. If the
Fermi momentum of the particle $x$ inside $a=(b+x)$ compensates for
the initial projectile velocity $v_a$, the low energy reaction
$A+x=B+c$ is induced at very low (even vanishing) relative energy
between $A$ and $x$. Basically, this technique extends the method of
transfer reactions to continuum states. Very successful results
using this technique have been reported \cite{Con07,Piz11}.

Another transfer method, coined as 
ANC technique \cite{Muk90,Tri06,Mu10} relies on fact that the amplitude for
the radiative capture cross section $b+x\longrightarrow a+ \gamma$
is given by $$M=<I_{bx}^a({\bf r_{bx}})|{\cal O}({\bf r_{bx}})|
\psi_i^{(+)}({\bf r_{bx}})>,$$ where $$I_{bx}^a=<\phi_a(\xi_b, \ \xi_x,\ {\bf %
r_{bx}}) |\phi_x(\xi_x)\phi_b(\xi_b)>$$ is the integration over the
internal coordinates $\xi_b$, and $\xi_x$, of $b$ and $x$,
respectively. For low energies, the overlap integral $I_{bx}^a$ is
dominated by contributions from large $r_{bx}$. Thus, what matters
for the calculation of the matrix element $M$ is the asymptotic
value of $I_{bx}^a\sim C_{bx}^a \ W_{-\eta_a, 1/2}(2\kappa_{bx}
r_{bx})/r_{bx}$, where $C_{bx}^a$ is the ANC and $W$ is the
Whittaker function. This coefficient is the product of the
{\it spectroscopic factor} and a normalization constant which depends on
the details of the wave function in the interior part of the
potential. Thus, $C_{bx}^a$ is the only unknown factor needed to
calculate the direct capture cross section. These normalization
coefficients can be found from: 1) analysis of classical nuclear
reactions such as elastic scattering [by extrapolation of the
experimental scattering phase shifts to the bound state pole in the
energy plane], or 2) peripheral transfer reactions whose amplitudes
contain the same overlap function as the amplitude of the
corresponding astrophysical radiative capture cross section. 
One of the many advantages of using transfer reaction techniques
over direct measurements is to avoid the treatment of the {\it electron screening
problem} \cite{Con07,Mu10}.

{\bf Intermediate energy Coulomb excitation} -  The Coulomb excitation cross section is given by
\begin{eqnarray}
{\frac{d\sigma_{i\rightarrow f}}{d\Omega}}&=&\left(
\frac{d\sigma}{d\Omega }\right)
_{\mathrm{el}}\frac{16\pi^{2}Z_{2}^{2}e^{2}}{\hbar^{2}}\nonumber \\
&\times& \sum
_{\pi\lambda\mu}{\frac{B(\pi\lambda,I_{i}\rightarrow
I_{f})}{(2\lambda
+1)^{3}}}\mid S(\pi\lambda,\mu)\mid^{2},\label{cross_2}%
\end{eqnarray}
where $B(\pi\lambda,I_{i}\rightarrow I_{f})$ is the reduced
transition probability of the projectile nucleus, $\pi\lambda=E1,\
E2,$ $M1,\ldots$ is the multipolarity of the excitation, and
$\mu=-\lambda,-\lambda+1,\ldots,\lambda$.

The relativistic corrections
to the Rutherford formula for $\left(
d\sigma /d\Omega\right)  _{\mathrm{el}}$ (relevant for collisions at  $50$ MeV/nucleon and above) has been investigated
in Ref. \cite{AAB90}.
It was shown that
the scattering angle increases by up to 6\% when relativistic
corrections are included in nuclear
collisions at 100 MeV/nucleon. The effect on the elastic scattering
cross section is even more drastic: up to $13\%$ for center-of-mass
scattering angles around 0-4 degrees.

The orbital integrals
$S(\pi\lambda,\mu$) contain the information about relativistic
corrections. Inclusion of absorption effects in
$S(\pi\lambda,\mu$) due to the imaginary part of an optical
nucleus-nucleus potential where worked out in Ref. \cite{BN93}.
These orbital integrals depend on the Lorentz factor
$\gamma=(1-v^{2}/c^{2})^{-1/2}$, with $c$ being the speed of light,
on the multipolarity $\pi\lambda\mu$, and on the {\it adiabaticity
parameter} $\xi (b)=\omega_{fi}b/\gamma v<1$, where
$\omega_{fi}=\left( E_{f}-E_{i}\right) /\hbar$ is the excitation
energy (in units of $\hbar$) and $b$ is the impact parameter.

Ref. \cite{Ber03} has shown that at 10 MeV/nucleon the
relativistic corrections are important only at the level of 1\%. At
500 MeV/nucleon, the correct treatment of the recoil corrections  is
relevant on the level of 1\%. Thus the non-relativistic treatment of
Coulomb excitation~\cite{AW75} can be safely used for energies below
about 10 MeV/nucleon and the relativistic treatment with a
straight-line trajectory~\cite{WA79} is adequate above about 500
MeV/nucleon. However at energies around 50 to 100 MeV/nucleon,
accelerator energies common to most radioactive beam facilities, it is very important to use a correct
treatment of {\it recoil and relativistic effects}, both kinematically and
dynamically.
At these energies, the corrections can add up to 50\%.
These effects were also shown in Ref.~\cite{AB89} for the case of
excitation of giant resonances in collisions at intermediate
energies.

A reliable extraction of useful nuclear properties,
like the electromagnetic response (B(E2)-values, $\gamma$-ray
angular distribution, etc.) from Coulomb excitation experiments at
intermediate energies requires a proper treatment of special
relativity \cite{Ber03,BCG03}. The dynamical relativistic effects have often been
neglected in the analysis of experiments elsewhere (see, e.g. \cite{Glas01}).
The effect is highly non-linear, i.e. a 10\% increase in
the velocity might lead to a 50\% increase (or decrease) of certain
physical observables. A general review of the importance of the relativistic
dynamical effects in intermediate energy collisions has been presented in
Ref. \cite{Ber05,Ber05_work}.

{\bf The Coulomb dissociation method} - The Coulomb dissociation method is quite simple.
The (differential, or angle integrated) Coulomb breakup cross section
for $a+A\longrightarrow b+c+A$ follows from Eq. \ref{cross_2}.
It can be rewritten as
\begin{equation}
{d\sigma_{C}^{\pi\lambda
}(\omega)\over d\Omega}=N^{\pi\lambda}(\omega;\theta;\phi)\ .\
\sigma_{\gamma+a\ \rightarrow\ b+c}^{\pi\lambda}(\omega),\label{CDmeth}
\end{equation}
where $\omega$ is the energy transferred from the relative motion to the
breakup, and $\sigma_{\gamma+a\ \rightarrow\ b+c}^{\pi\lambda}(\omega)$ is the photo nuclear cross
section for the multipolarity ${\pi\lambda}$ and photon energy $\omega$. The
function $N^{\pi\lambda}$, sometimes called {\it virtual photon numbers}, depends on $\omega$, the relative motion energy,
nuclear charges and radii, and the scattering angle $\Omega=(\theta,\phi)$.
$N^{\pi\lambda}$
can be reliably calculated \cite{Ber88} for each
multipolarity ${\pi\lambda}$. Time reversal allows one to deduce the radiative
capture cross section $b+c\longrightarrow a+\gamma$ from $\sigma_{\gamma+a\ \rightarrow\ b+c}%
^{\pi\lambda}(\omega)$. This method was proposed in Ref. \cite{BBR86} and has
been tested successfully in a number of reactions of interest for astrophysics.
The most celebrated case is the
reaction $^{7}$Be$($p$,\gamma)^{8}$B \cite{Tohru}, followed by numerous
experiments in the last decade (see e.g. Ref. \cite{EBS05}).

Eq. \ref{CDmeth} is based on first-order perturbation theory. It
also assumes that the nuclear contribution to the breakup is small,
or that it can be separated under certain experimental conditions.
The contribution of the nuclear breakup has been examined by several
authors (see, e.g. \cite{BG98}). $^8$B has a small proton separation
energy ($\approx 140$ keV). For such loosely-bound systems it had
been shown that multiple-step, or higher-order effects, are
important \cite{BC92}. These effects occur by means of
continuum-continuum transitions. Detailed studies of dynamic
contributions to the breakup were explored in refs.
\cite{BBK92,BB93} and in several other publications which followed.
The role of higher multipolarities (e.g., E2 contributions
\cite{Ber94,GB95,EB96} in the reaction $^{7}$Be$($p$,\gamma)^{8}$B)
and the coupling to high-lying states has also to be investigated
carefully. It has also been shown that the
influence of giant resonance states is small \cite{Ber02}. 

{\bf Knock-out reactions} - The early interest in knockout reactions came from studies of
nuclear halo states, for which the narrow momentum distributions
of the core fragments
in a qualitative way revealed the large spatial extension of the
halo wave function. It was shown 
\cite{ber92} that the longitudinal component of the momentum
(taken along the beam or $z$ direction) gave the most accurate
information on the intrinsic properties of the halo and that it
was insensitive to details of the collision and the size of the
target. In contrast to this, the transverse distributions of the
core are significantly broadened by diffractive effects and by
Coulomb scattering. For experiments that observe the nucleon
produced in elastic breakup, the transverse momentum is entirely
dominated by diffractive effects, as illustrated \cite{ann94} by
the angular distribution of the neutrons from the reaction
$^{9}$Be($^{11}$Be,$^{10}$Be+n)X. In this case, the width of the
transverse {\it momentum distribution} reflects essentially the size of
the target  \cite{BH04}.

\begin{figure}[h]
\includegraphics[width=13.5pc]{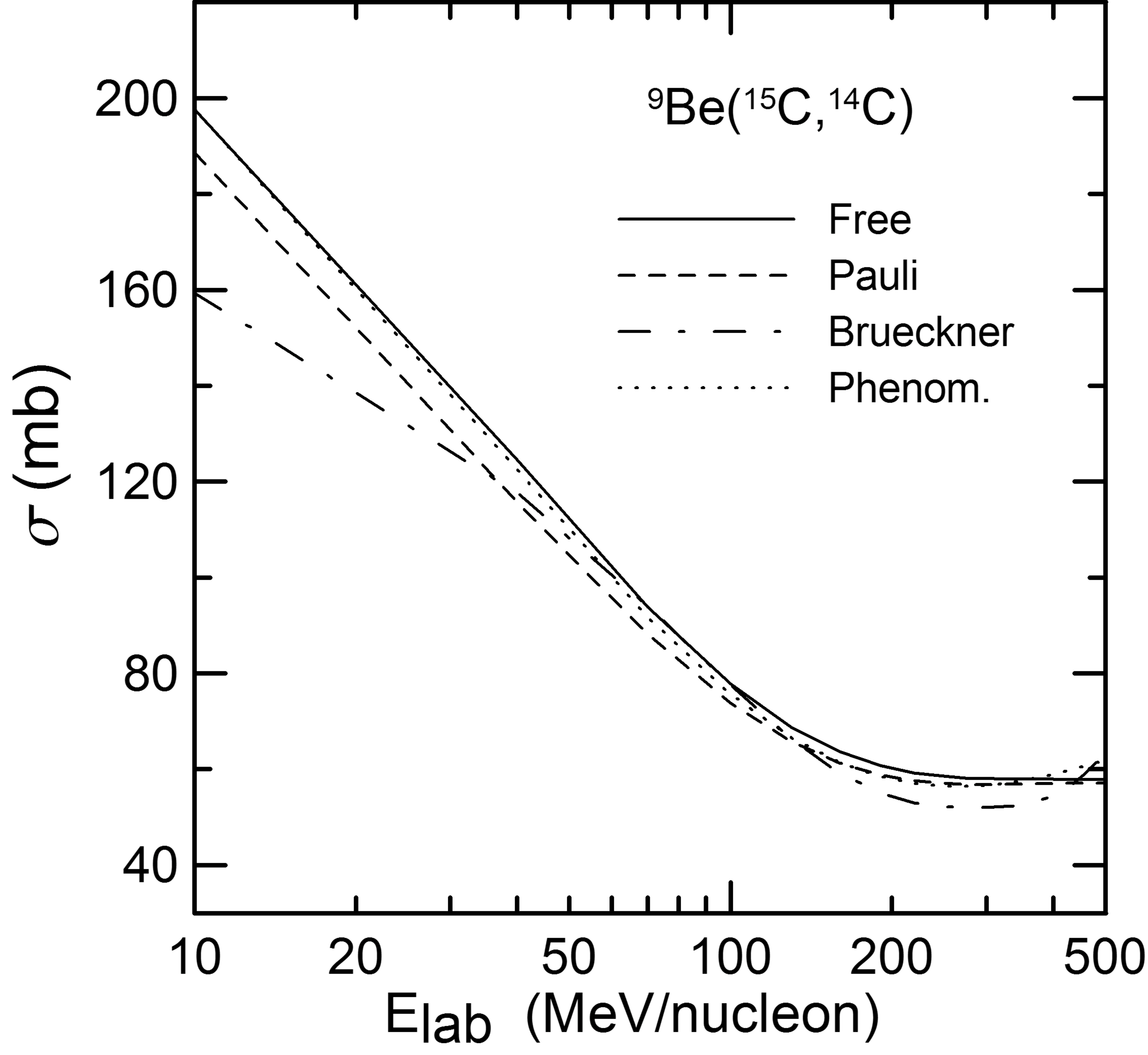}\hspace{2pc}%
\begin{minipage}[b]{14pc}\caption{\label{sko15c}Total  knockout cross sections for removing the $l=0$ halo neutron of $^{15}$C, bound by 1.218 MeV, in the reaction $^9$Be($^{15}$C,$^{14}$C$_{gs}$). The solid curve is obtained with the use of free nucleon-nucleon cross sections.  The dashed curve includes the geometrical effects of Pauli blocking. The dashed-dotted curve is the result using the Brueckner theory,  and the dotted curve is a  phenomenological parametrization.}
\end{minipage}
\end{figure}

To test the influence of the medium effects in nucleon knockout reactions, we consider the removal of the $l=0$ halo neutron of $^{15}$C, bound by 1.218 MeV. The reaction studied is $^9$Be($^{15}$C,$^{14}$C$_{gs}$). The total cross sections as a function of the bombarding energy are shown in figure \ref{sko15c}. The solid curve is obtained with the use of free nucleon-nucleon cross sections.  The dashed curve includes the geometrical effects of Pauli blocking. The dashed-dotted curve is the result using the {\it Brueckner theory},  and the dotted curve is the phenomenological parametrization of the free cross section.

\begin{figure}[h]
\includegraphics[width=13.5pc]{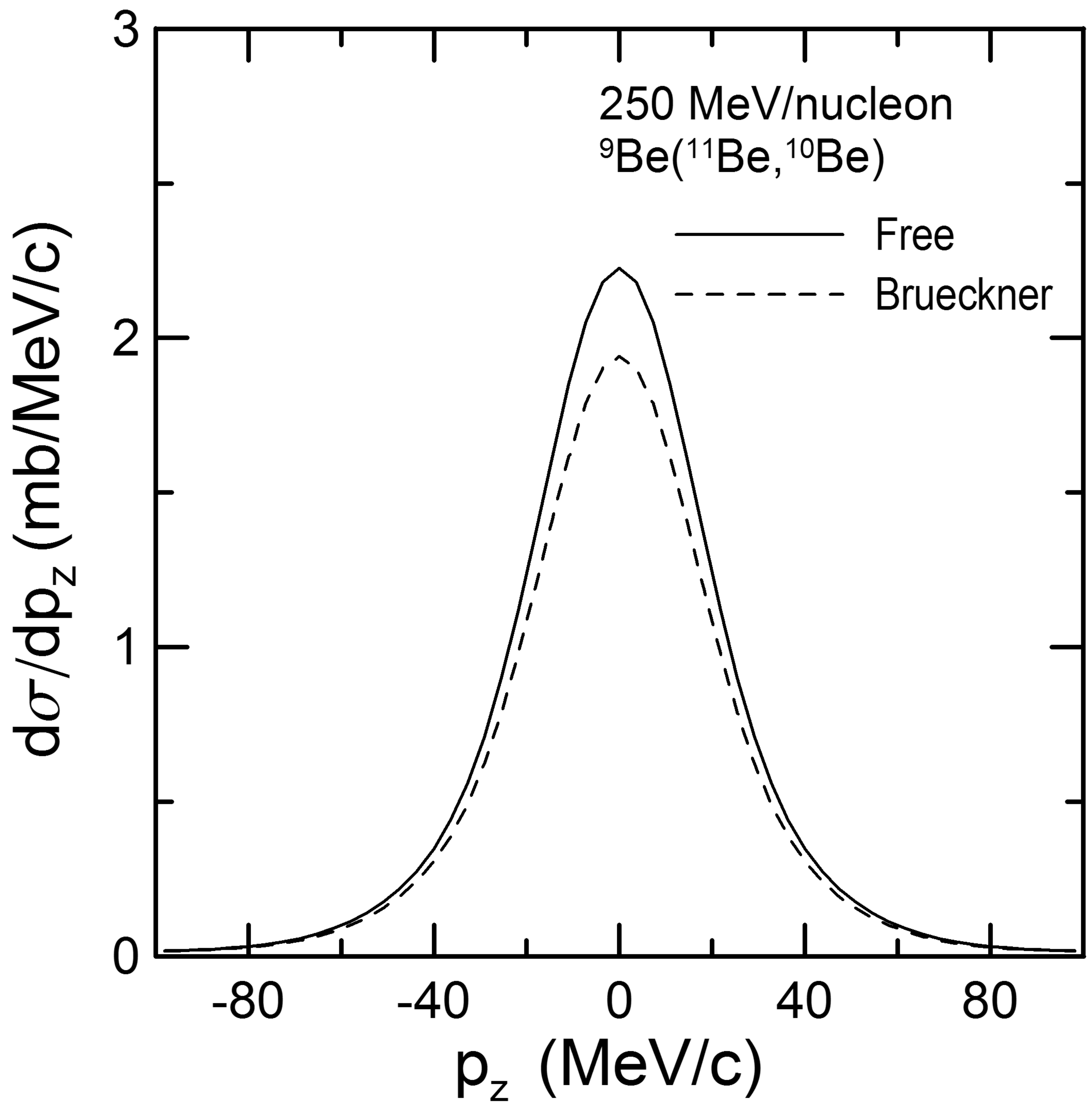}\hspace{2pc}%
\begin{minipage}[b]{14pc}\caption{\label{dsdpz}Longitudinal momentum distribution for the residue in
the $^9$Be($^{11}$Be,$^{10}$Be),  reaction at 250 MeV/nucleon. The dashed curve is the cross
section calculated using the NN cross section from the Brueckner theory and the solid curve is obtained  the free cross section.}
\end{minipage}
\end{figure}

In figure \ref{dsdpz}  we plot the longitudinal momentum distributions for the reaction $^9$Be($^{11}$Be,$^{10}$Be),  at 250 MeV/nucleon \cite{BC10}. The dashed curve is the cross section calculated using the NN cross section from the Brueckner theory and the solid curve is obtained  the free cross section. One sees  that the momentum distributions are reduced by 10\%, about the same as  the total cross sections, but the shape remains basically unaltered. If one rescales the dashed curve to match the solid one, the differences in the width are not visible \cite{Kar11}.

\section{Conclusions}
There were many questions not addressed in this review, such as the role of central nucleus-nucleus collisions in determining phase transition, equation of state, and a quark-gluon plasma, all topics or relevance in astrophysics. The review was more focused on the role of short-lived, exotic nuclei. There are many important scientific questions to be addressed both experimentally and theoretically in nuclear physics of exotic nuclei with relevance for astrophysics.
These questions provide extreme challenges for experiments and theory. On the experimental side, producing the beams of radioactive nuclei needed to address the scientific questions has been an enormous challenge. Pioneering experiments have established the techniques and present-generation facilities have produced first exciting science results, but the field is still at the beginning of an era of discovery and exploration that will be fully underway once the range of next- generation facilities becomes operational. The theoretical challenges relate to wide variations in nuclear composition and rearrangements of the bound and continuum structure, sometimes involving near-degeneracy of the bound and continuum states. The extraction of reliable information from experiments requires a solid understanding of the reaction process, in addition to the structure of the nucleus. In astrophysics, new observations, for example the expected onset of data on stellar abundances, will require rare-isotope science for their interpretation.

\bigskip

The author acknowledge financial support from  the US Department of Energy Grants DE-FG02- 08ER41533, DE-SC0004971.

\end{document}